# *Theoretical derivation of blood velocity from TOF-MRA based artery centerline*

**Abrar Faiyaz** *

Department of Neurology, University of Rochester, Rochester, NY 14609, USA;
* Correspondence: abrar_faiyaz@urmc.rochester.edu; Tel.: +1-(585)+355-9186

**Abstract**

Time-of-flight magnetic resonance angiography (TOF-MRA) is widely used for noninvasive visualization of arterial anatomy, but extracting hemodynamics like blood velocity typically requires supplementary phase-contrast scans, tagging or multi-TE images. This study proposes a novel, physics-informed computational framework to extract variable fluid velocity directly from standard TOF-MRA signal profiles. We analytically expand the approach-to-steady-state Bloch equations to include convective flow, establishing a mathematical relationship between the spatial decay of longitudinal magnetization and fluid velocity. The velocity derivation was further extended to pointwise estimation over a 1-D centerline, overcoming the limitations of constant-velocity assumptions. To validate and solve this problem, a MATLAB (R2025b) simulation framework was developed to model fluid flow in two variable-geometry flowing tube cases, i.e., continuous narrowing and focal stenosis, under synthetic scanner noise. A global inverse optimization approach utilizing Dual-Tikhonov regularization was applied to stably invert the ill-posed transit time integral, actively penalizing high-frequency numerical ringing while preserving structural curves. The computational simulations successfully recovered ground-truth point-wise velocities, tracking gradual hemodynamic accelerations and sharp stenotic jets. This theoretical framework and the example centerline TOF-MRA signal intensity provide a robust mathematical proof-of-concept that quantitative, localized functional hemodynamic metrics can be extracted from standard structural MRA imaging, establishing a foundation for advanced flow quantification without requiring additional scan time.

## 1. Introduction

To maintain metabolic and neural function, the human brain requires continuous and tightly regulated blood flow. Consequently, the quantitative assessment of cerebrovascular hemodynamics is important for the diagnosis and longitudinal management of neurovascular pathologies, including cerebral small vessel disease (CSVD) [1,2]. In standard clinical cases, structural and functional evaluations of the cerebrovascular network are typically segregated across individual imaging modalities [3]. Three-dimensional time-of-flight magnetic resonance angiography (TOF-MRA) is a primary non-invasive modality for high-resolution structural imaging of the arteries. It effectively delineates the anatomical topology of the intracranial arterial network without requiring exogenous contrast agents [4,5].

However, the clinical utility of TOF-MRA is mainly constrained to qualitative structural assessments of macroscopic arterial abnormalities [4,6]. As the hemodynamic metrics are not usually acquired and processed in a short time for critical clinical cases, for example, in cases like stroke and emergency scenarios, only structural arterial imaging is prioritized. This diagnostic paradigm frequently overlooks the quantitative hemodynamic metrics that are essential for characterizing micro- and macrovascular alterations and pathogenesis investigations. Currently, the extraction of functional hemodynamics, such as absolute fluid velocity, necessitates supplementary acquisitions, including phase-contrast MRI (PC-MRI) or transcranial Doppler (TCD) ultrasound [7]. The reliance on multiple, independent imaging sequences inherently increases total acquisition time, introduces cross-modality registration challenges, and elevates healthcare costs. Another approach that proposed multi-TE acquisition for quantitative velocity, like TOF-MRA, also comes with the burden of acquisition protocol being inherently different than clinical TOF-MRA [6,8].

To address these limitations, this study introduces a physics-informed computational framework designed to obtain fluid velocity directly from standard TOF-MRA signal centerline intensity profiles. By analytically applying the Bloch equations to model spatial signal decay within simulated flowing tubes, this approach establishes a mathematical foundation for continuous flow quantification. Extracting these functional metrics from the structural topology provides a unified diagnostic pipeline, enabling concurrent anatomical and hemodynamic assessment without requiring any additional scan time. It further posits the potential to be applied to already scanned TOF-MRA scans.

## 2. Materials and Methods

The proposed theoretical framework analytically expands the Bloch equations to establish a mathematical relationship between fluid velocity and the spatial decay of prepulse longitudinal magnetization. TOF-MRA relies heavily on the entry slice phenomenon (or flow-related enhancement), where fully relaxed blood entering the imaging volume exhibits maximum signal intensity due to its lack of prior radiofrequency saturation. As blood flows along a vessel's centerline (spatial coordinate z) at a mean velocity v, it experiences consecutive radiofrequency pulses separated by the repetition time TR. Assuming the arterial segment resides within a single slab, this progressive saturation characterizes the blood's approach to steady state.

### *2.1. Deriving Effective Relaxation Rate for Blood*

When fresh blood enters the imaging volume at $z = 0$, its longitudinal magnetization is completely relaxed at thermal equilibrium, meaning $M_z(0) = M_0$. After absorbing an integer history of n successive radiofrequency pulses, the moving packet's transient approach toward the steady state is described by [9]

$$M_z(n) = M_{SS} + (M_0 - M_{ss})(E_1 \cos\alpha)^n \tag{1}$$

where $E_1 = \exp(-TR/T1)$ represents the $T1$ relaxation factor, and the Ernst steady-state longitudinal magnetization is defined as

$$M_{ss} = M_0 + \frac{1 - E_1}{1 - E_1 \cos\alpha} \tag{2}$$

with $\alpha$ representing the sequence's flip angle.

Because the blood packet with T1 travels at velocity v(z), the distance z traversed is proportional to the elapsed time. Therefore, the number of accumulated pulses n upon reaching coordinate z is continuously approximated by n = $\frac{z}{v \cdot TR}$. Substituting this spatial mapping transforms the time-domain evolution into a purely spatial decay profile:

$$M_z(z) = M_{ss} + (M_0 - M_{ss})(e^{-\frac{TR}{T1}} \cos\alpha)^n \tag{3}$$

We can rewrite the exponential term to the following using an $x^y = e^{y \ln x}$ identity,

$$(e^{-\frac{TR}{T1}} \cos\alpha)^n = (e^{-\frac{TR}{T1}} \cos\alpha)^{\frac{z}{v \cdot TR}}$$

$$= \exp\left(\frac{z}{v \cdot TR}\left(\ln(\cos\alpha) - \frac{TR}{T1}\right)\right)$$

$$= \exp\left(-\frac{z}{v \cdot \mathrm{TR}}\left(-\ln(\cos\alpha) + \frac{TR}{T1}\right)\right)$$

To enable non-linear curve fitting, the exponential term is analytically combined to define a novel physical parameter for blood, i.e., the effective relaxation rate $R_{eff}$,

$$R_{eff} = \frac{1}{T1} - \frac{\ln(\cos\alpha)}{TR} \tag{4}$$

(Note: Because if $\cos\alpha < 1$, the natural log $\ln(\cos\alpha)$ is negative. Therefore, both terms contribute positively to the saturation rate. The Radiofrequency (RF) pulses act as an artificial, accelerated "relaxation" driving the signal down.)

The equation simplifies to

$$M_z(z) = M_{ss} + (M_0 - M_{ss}) \exp\left(-\frac{z}{v}\left(R_{eff}\right)\right) \tag{5}$$

The signal intensity measured by the scanner is proportional to the transverse magnetization after the RF pulse,

$$S(z) \propto M_z(z) \sin\alpha \tag{6}$$

So,

$$S(z) = S_{ss} + (S_0 - S_{ss}) \exp\left(-\frac{R_{eff}}{v} z\right) \tag{7}$$

By extracting the spatial intensity profile from a TOF-MRA scan $S(z), S_{ss}$ and $S_0$, non-linear exponential regression can solve for the empirical decay constant $k_{eff}$.

So,

$$k_{eff} = \frac{R_{eff}}{v} \tag{8}$$

This allows the blood velocity $v$ to be theoretically inverted using the known scanner parameters.

*2.2. Deriving Position-Specific Velocity Along Artery Centerline*

If velocity is a 1D array varying along the centerline from starting landmark to ending landmark, v(z), the time it takes for a packet of blood to reach distance z is no longer a simple division. It is the cumulative sum of the time spent in each infinitesimally small segment of the tube. Mathematically, this is an integral

$$t(z) = \int_0^z \frac{1}{v(z')} dz' \tag{9}$$

Because the blood is experiencing RF pulses over this entire transmit time, the standard exponential decay equation changes from a scalar multiplication to an integral of the local decay-rate.

$$S(z) = S_{ss} + (S_0 - S_{ss}) \exp\left(-R_{eff} \int_0^z \frac{1}{v(z')} dz'\right) \tag{10}$$

The signal decay at any point $z$ now becomes a function of the entire velocity history of the blood packet from the starting landmark up to that point.

*2.3. Estimation of v(z) on Centerline*

Reorganizing and taking natural logarithm on both sides of the signal, from Equation (10),

$$ln\left(\frac{S(z) - S_{ss}}{(S_0 - S_{ss})}\right) = -R_{eff}\int_0^z \frac{1}{v(z')}dz' \quad (11)$$

By applying the Fundamental Theorem of Calculus to the right side, the derivative cancels the integral,

$$\frac{d}{dz}\left[ln\left(\frac{S(z) - S_{ss}}{(S_0 - S_{ss})}\right)\right] = -\frac{R_{eff}}{v(z)} \quad (12)$$

Using chain rule on the left side, the equation simplifies to

$$\left(\frac{S'(z)}{S(z) - S_{ss}}\right) = -\frac{R_{eff}}{v(z)} \quad (13)$$

$$v(z) = -R_{eff}\left(\frac{S(z) - S_{ss}}{S'(z)}\right) \quad (14)$$

*2.4. Computational Simulation of Realistic Blood Flow Scenarios in 2D*

Because the mathematical derivation isolates the 1D centerline intensity decay, this framework is inherently applicable to multi-dimensional (2D and 3D) TOF-MRA acquisitions once the vascular graph is isolated. To validate the derivation against ground-truth hemodynamics, a computational MATLAB framework was implemented to model fluid flow under standard clinical 3T scanner sequence parameters (TR = 30 ms, flip angle $\alpha$ = 25°, arterial longitudinal relaxation time T1 = 1.6 s, with added synthetic scanner noise).

Flow was simulated in two distinct geometric flowing tube cases representing a realistic healthy and a pathological morphology. The initial proximal radius was set to 0.15 cm with a starting inflow velocity of 20 cm/s. Ground-truth variable velocity profiles, v(z), were established using the principle of mass conservation, dictating that fluid velocity is inversely proportional to the cross-sectional area ($v \propto \frac{1}{r^2}$).

- Scenario A (Distal Narrowing): Modeled a healthy vessel that gradually narrows to a distal radius of 0.10 cm, simulating natural anatomical narrowing and a corresponding smooth hemodynamic acceleration.
- Scenario B (Focal Stenosis): Modeled localized pathological narrowing, applying an inverted Gaussian reduction in radius down to a focal minimum of 0.08 cm at the middle of the tube, creating a sharp, high-velocity stenotic jet.

At first, a constant velocity model was globally fit to both scenarios to show standard assumptions, and then variable velocity model was fit to report point-wise velocity at the centerline.

*2.5. Global Regularized Inverse Optimization via Dual-Tikhonov Regularization*

While sequential spatial differentiation effectively isolates the centerline decay, standard numerical derivatives are highly susceptible to cascading high-frequency noise amplification, particularly in regions of slow signal evolution. To systematically eliminate these numerical singularities and prevent oscillatory ringing artifacts in the final velocity profile, the inverse problem was formulated as a single global matrix optimization using Dual-Tikhonov regularization.

By spatially upsampling the centerline to a high-density computational grid, the Bloch equation transit time integral can be mapped as a linear system

$$Lu \approx y \quad (15)$$

Here, $y$ represents the discrete spatial integral derived from the log transformed normalized MRA signal, $u$ represents the unknown array of inverse velocities (1/v), and L is a lower triangular cumulative integration matrix scaled by the spatial step dimension $\Delta z$.

$$y = -\frac{1}{R_{eff}} ln\left(\frac{S - S_{ss}}{(S_0 - S_{ss})}\right) \tag{16}$$

To stably invert this ill-posed system, a dual-penalty cost function was deployed to simultaneously enforce data fidelity and structural smoothness. The objective function minimizes the global error while applying regularization matrices for both the first derivative (D1) and the second derivative (D2),

$$\arg\min_u \left( ||Lu - y||_2^2 + \lambda_1 ||D_1 u||_2^2 + \lambda_2 ||D_2 u||_2^2 \right) \tag{17}$$

The first-derivative penalty ($\lambda$1) strictly suppresses localized high-frequency noise spikes (jaggedness), while the second-derivative penalty ($\lambda$2) controls the broader geometric curvature (stiffness) of the profile. Because this formulation establishes a globally convex quadratic problem, it bypasses non-linear optimization completely, yielding a direct analytical solution

$$u = \left(L^T L + \lambda_1 {D_1}^T D_1 + \lambda_2 {D_2}^T D_2\right)^{-1} L^T y \tag{18}$$

Following the matrix inversion, the array u is inverted pointwise to reconstruct the continuous localized velocity profile v(z), bounded by a maximum physiological limit of 150 cm/s to ensure stability.

## 3. Results

### *3.1. Forward Simulation of Hemodynamics in Flowing Tubes*

The forward computational model successfully mapped the 1D theoretical velocity profiles into 2D spatial flow fields (Figure 1). By applying a parabolic laminar flow constraint within the synthetic tissue background, the simulations visually confirmed the principle of mass conservation across both morphologies. Scenario A (Narrowing Tube) demonstrated a progressive and smooth acceleration of the central fluid column as the cross-sectional area decreased. Scenario B (Stenotic Tube) accurately localized a high-velocity jet exactly at the z = 6 cm focal narrowing, with the corresponding quiver vectors indicating an increase in axial momentum strictly bounded by the vessel walls.

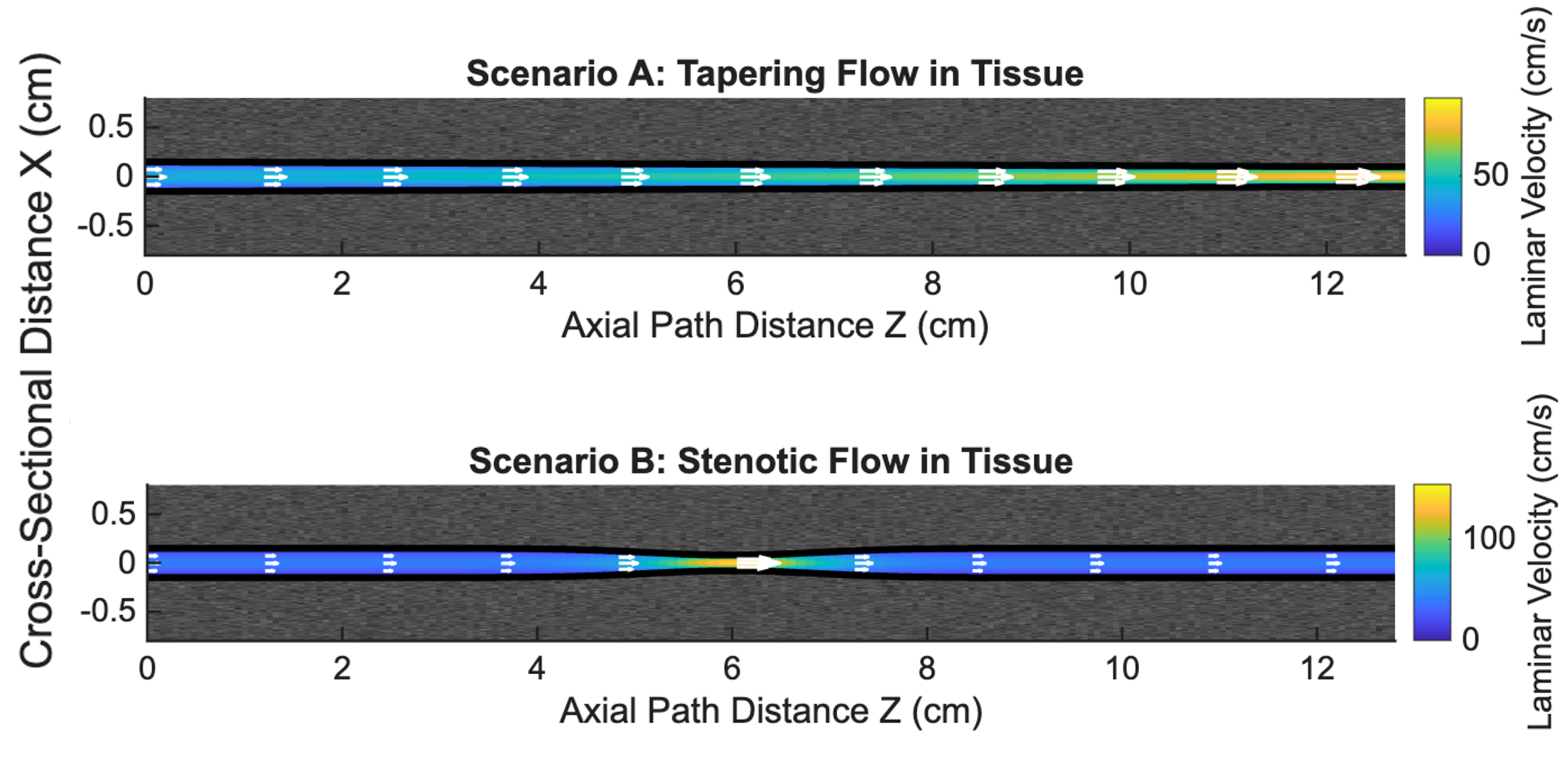

**Figure 1.** Scenario (**A**) Narrowing/Tapering Flow; (**B**) Stenotic Flow; velocity field shown in perula colormap for both cases. Simulation conducted for an assumed artery segment of 12 cm. The arrows indicated direction of blood flow.

### *3.2. Inverse Estimation and Regularization Sensitivity*

Application of the standard global constant velocity assumption failed to capture the localized hemodynamic changes in either simulated morphology, yielding flat mean estimates (18.8 cm/s for Scenario A; 20.4 cm/s for Scenario B) that significantly deviated from the true variable profiles (Figure 2, top row).

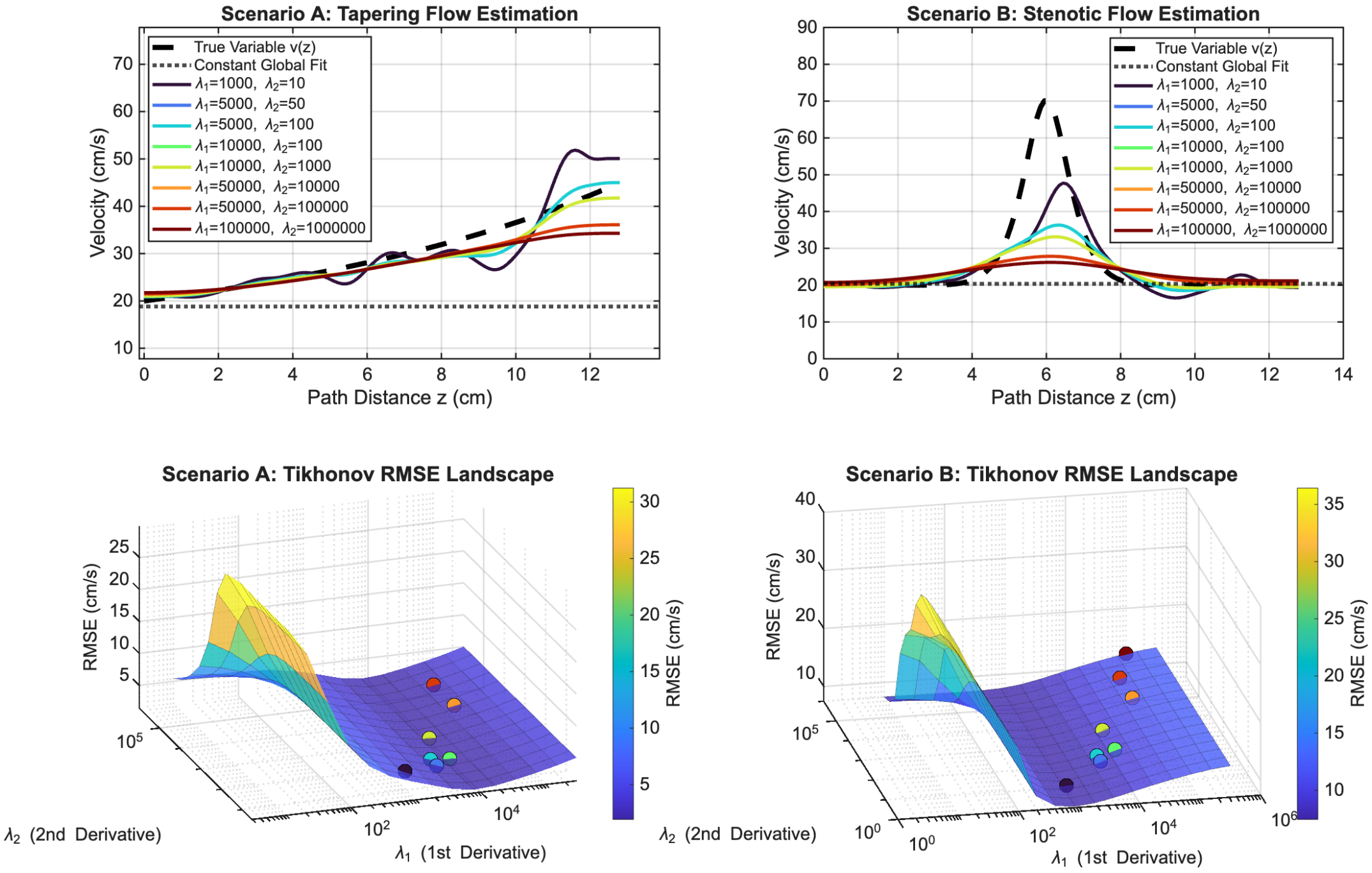

**Figure 2.** Evaluation of Dual-Tikhonov Regularization for Variable Velocity Estimation. (**Top Row**) Reconstructed spatial velocity profiles for the continuous tapering tube (Scenario (**A**), left) and the focal stenotic tube (Scenario (**B**), right). The estimations are derived across a defined sweep of first derivative ($\lambda 1$) and second-derivative ($\lambda 2$) penalty weights and compared against both the theoretical ground-truth variable velocity (thick dashed line) and the standard constant velocity assumption (dotted line). (**Bottom Row**) 3D root-mean-square error (RMSE) landscapes mapping the estimation accuracy across the continuous regularization parameter space for each scenario. The colored markers represent the specific discrete parameter sets evaluated in the top row, illustrating the optimization convergence minimum and highlighting the inherent L2-norm trade-off between baseline noise suppression and the preservation of sharp hemodynamic peaks. Note: 2/3 lines shown in the legend are overlapped by other lines, obscuring their view.

The proposed Dual-Tikhonov inverse solver successfully reconstructed the point-wise variable velocity, v(z), directly from the simulated noisy MRA signals. The parameter sweep evaluated the mathematical trade-off between first-derivative noise suppression ($\lambda 1$) and second-derivative structural stiffness ($\lambda 2$) across the scenarios, utilizing a discrete set of regularization parameters (Table 1) chosen empirically to span a broad range of penalty magnitudes.

- Scenario A (Narrowing Tube): The estimation achieved high fidelity across most of the parameter space, with root-mean-square error (RMSE) ranging from 1.94 cm/s to 4.21 cm/s. Because the true velocity undergoes a gradual, low-frequency transition, higher regularization weights ($\lambda 1 \geq 10{,}000$) provided the lowest absolute error by heavily suppressing baseline scanner noise without distorting the underlying curve.
- Scenario B (Stenotic Tube): The inverse estimation successfully tracked the sharp spatial transient of the stenotic jet. RMSE ranged from 7.71 cm/s to 12.17 cm/s across the tested parameters. The lowest absolute RMSE (7.71 cm/s) was recorded at the weakest regularization bounds (λ1 = 1000, λ2 = 10); however, qualitative observation of the reconstructed profile revealed persistent high-frequency baseline oscillations. Moderate regularization weights (e.g., λ1 = 5000, λ2 = 100) established an optimal balance, effectively mitigating oscillatory ringing artifacts in the stable proximal and distal segments while preserving the magnitude of the 70 cm/s peak.

**Table 1.** Dual-Tikhonov Regularization Parameter Sets.

| Set | λ1 (1st Derivative) | λ2 (2nd Derivative) | Evaluated Effect |
|---|---|---|---|
| 1 | 1000 | 10 | Weak stabilization; persistent high-frequency ringing. |
| 2 | 5000 | 50 | Approaching structural stability. |
| 3 | 5000 | 100 | Optimal balance; maximal peak preservation with stable baseline. |
| 4 | 10,000 | 100 | Stronger jaggedness penalty. |
| 5 | 10,000 | 1000 | Initiation of artificial curvature stiffening. |
| 6 | 50,000 | 10,000 | Moderately over-smoothed profile. |
| 7 | 50,000 | 100,000 | Pronounced flattening of the stenotic jet. |
| 8 | 100,000 | 1,000,000 | Severely over-regularized; ignores underlying MRI data. |

The mapped 3D error landscapes (Figure 2, bottom row) confirm the presence of a distinct convergence minimum in the regularization space. Severe over-regularization (λ2 ≥ 100,000) predictably resulted in the highest error margins by mathematically flattening the velocity variations toward a constant mean, underscoring the necessity of properly tuned penalty matrices for accurate physiological modeling.

Current imaging modalities for assessing cerebrovascular hemodynamics often force a tradeoff between scan duration, invasiveness, and spatial context (Table-2). Techniques like 4D Phase Contrast MRI and Digital Subtraction Angiography (DSA) provide detailed velocity and flow dynamics. However, they require lengthy scan times of up to 20 min and 90 min respectively. They also introduce risks like velocity aliasing or ionizing radiation. Noninvasive alternatives like Transcranial Doppler (TCD) ultrasound and Arterial Spin Labeling (ASL) face different challenges. They are limited by operator dependency, poor acoustic windows, or an inability to measure conduit arterial velocity. In contrast, TOF-MRA is a widely utilized rapid sequence. It requires only 4 min to map arterial structural topology. Conventionally, this modality is limited to qualitative assessments. Extracting quantitative hemodynamic profiles directly from this signal data overcomes this limitation.

**Table 2.** Comparison of imaging modalities for artery imaging and their limitations.

| Imaging Modality | Primary Contrast Mechanism | Target Output | Average Scan time* | Limitations |
|---|---|---|---|---|
| TOF-MRA [6,10] | Flow-Related Enhancement (Amplitude) | Arterial Structural Topology | ~4 min | Conventionally limited to qualitative structural assessment. |
| PC-MRI/4D Flow [11–13] | Bipolar Gradient Phase Shifts | Multi-directional Velocity, Flow Volume | 5–20 min (4D Flow) | Long scan times, low resolution, VENC aliasing vulnerability. |
| TCD Ultrasound [14,15] | Acoustic Doppler Shift | Real-time Beat-to-Beat Velocity | 15–30 min | Operator dependent, poor acoustic windows, lacks spatial context. |
| ASL [16,17] | RF Magnetic Tagging of Blood Water | Tissue Perfusion, Arterial Transit Time | 4–8 min | Low SNR, measures capillary perfusion rather than conduit velocity/arterial property. |
| DSA (Optical Flow) [16,18,19] | Exogenous Radiocontrast Dynamics | High-Resolution Angiography, Relative Velocity | 30–90 min (Full procedure) | Invasive, utilizes ionizing radiation and iodinated contrast. |

* Please note that exact scan times can vary based on the specific clinical indication, the hardware being used (e.g., modern MRI accelerators), and the patient's condition.

To bridge the theoretical simulations with practical clinical implementation, Figure 3 demonstrates the successful extraction of 1D macroscopic signal intensities along the complex 3D centerlines of an in vivo human cerebrovascular network (data sourced from the TopBrain Challenge). The mapped spatial decay visually aligns with the progressive signal attenuation modeled in Scenario A, confirming the feasibility of evaluating generalized tapering tube geometries within real physiological datasets. Furthermore, the inclusion of the focal stenosis model (Scenario B) establishes a mathematical baseline for quantifying localized pathogenic narrowing, which is critical for evaluating microvascular anomalies.

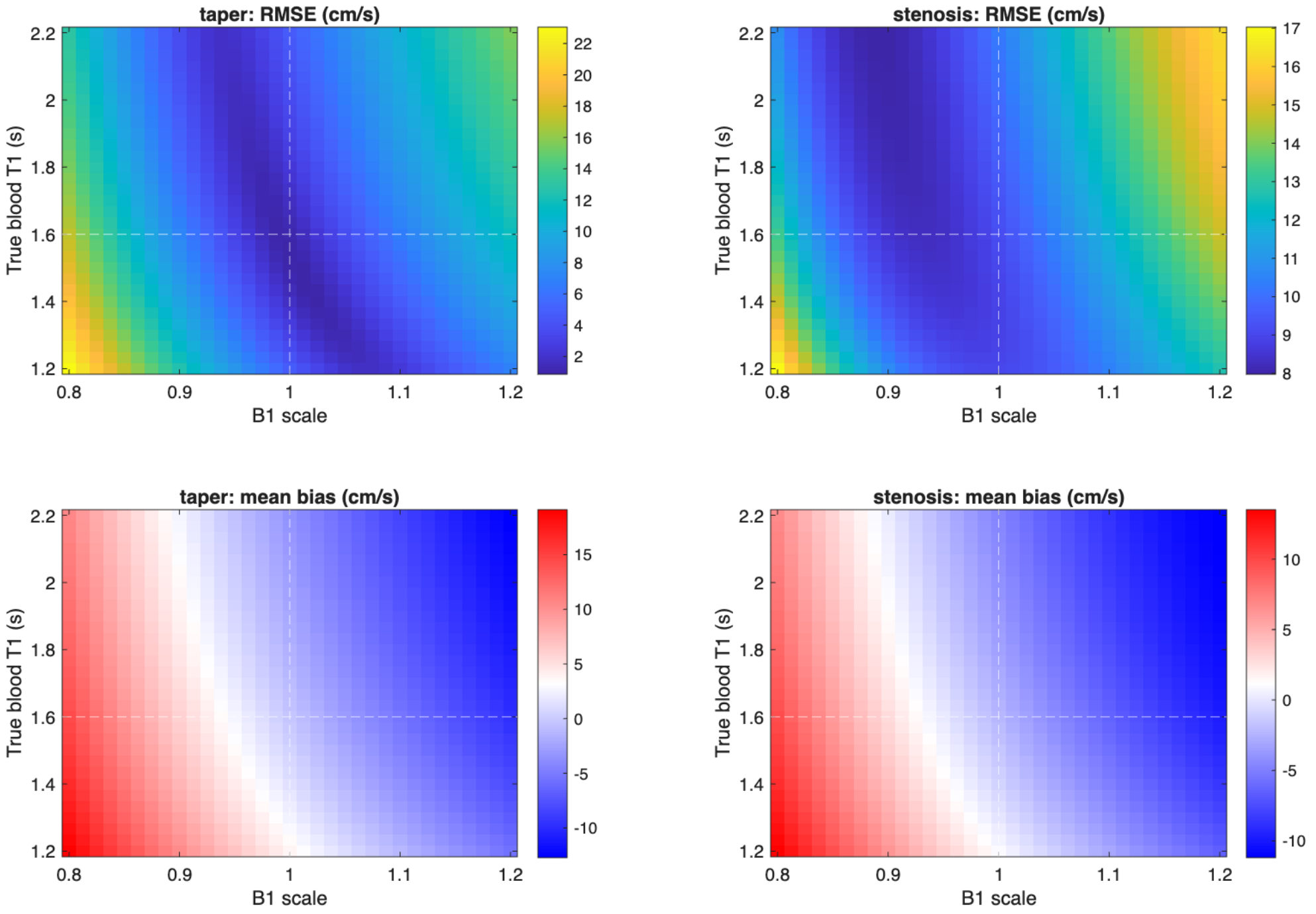

**Figure 3.** Sensitivity of TOF-MRA velocity estimates to blood T1 and B1+/flip-angle uncertainty. Heatmaps show the RMSE and mean bias of Tikhonov-regularized velocity estimates when synthetic TOF-MRA signals were generated with varied true blood T1 and B1 scale values, while inversion was performed using nominal parameters. Dashed white lines indicate the nominal assumptions used in the inverse model: blood T1 = 1.6 s and B1 scale = 1.0. RMSE maps quantify total velocity-estimation error, whereas bias maps show the signed mean deviation from the ground-truth velocity profile. Both tapering and focal stenosis scenarios demonstrate sensitivity to flip-angle scaling, with additional modulation by blood T1.

### *3.3. Sensitivity to Blood T1 and Flip Angle Uncertainty*

Figure 3 summarizes the sensitivity of the Tikhonov velocity estimates to mismatches in blood T1 and B1+ flip-angle scaling. Across the systematic perturbation grid, the inversion framework yielded a maximum RMSE of 20.7 cm/s and a maximum absolute mean bias of 16.6 cm/s for the distal tapering case, while the focal stenosis case demonstrated a maximum RMSE of 18.1 cm/s and a maximum absolute mean bias of 11.0 cm/s. Analysis of the signed bias maps revealed a distinct dependence on flip-angle scaling, where a lower assumed effective excitation resulted in a positive velocity bias and a higher flip-angle scaling led to a negative bias. Variations in T1 further modulated this effect, with the most significant errors occurring when deviations in both T1 and the flip angle were compounded. Notably, even near the nominal parameter baseline, the focal stenosis case maintained a higher baseline RMSE, reflecting the inherent mathematical challenge of accurately recovering a sharply varying, high-gradient velocity profile under regularized inversion.

## 4. Discussion

The results of this study validate a physics-informed computational framework capable of extracting quantitative variable hemodynamics directly from TOF-MRA. By analytically expanding the approach-to-steady-state Bloch equations to incorporate convective flow properties, a mathematical relationship linking spatial signal decay to -blood velocity was established. While traditional methods rely on multi-echo or phase contrast-based approaches, this study demonstrates the capacity to reconstruct point-wise variable velocity profiles in both tapering and stenotic flowing tubes. The proposed Dual-Tikhonov regularization framework recovered ground-truth variable velocities with RMSE of 1.94 cm/s for continuous tapering morphologies and 7.71 cm/s for sharp focal stenoses. By extracting these metrics without requiring supplementary functional sequences like phase-contrast MRI, this framework offers a computationally efficient approach to streamline the assessment of cerebrovascular conditions.

Initially theoretical iterations were limited by relying on the fundamental assumption that blood flow velocity (v) remains constant along the fitted 3D centerline (Section 2.1). In physiological environments, hemodynamics exhibits complex spatial variations due to changing cross-sectional areas. Assuming a constant $v$ mathematically simplifies the pulse history term ($n = v \cdot TRz$), but inherently masks focal hemodynamic alterations, yielding only an "effective" average velocity over the extracted segment. By transitioning to a continuous integral formulation for the transit time ($n = \int v(z) \cdot TR1\, dz$) and deploying a global matrix optimization, this study successfully isolated local velocity fluctuations, overcoming the numerical instability of piecewise spatial approximations. (Sections 2.2 and 2.3)

The accuracy of this variable inverse solver is demonstrably dependent on the applied regularization parameters. The simulation results underscore the necessity of a dual-penalty system to prevent the mathematical singularities typically associated with differentiating noisy MRA signals. As mapped in the error landscapes, optimizing both the first derivative ($\lambda 1$ = 5000) and second derivative ($\lambda 2$ = 100) penalties effectively mitigated high-frequency oscillatory ringing while preserving the maximum amplitude of the stenotic jet.

Conversely, over-regularizing the curvature penalty artificially flattened true hemodynamic peaks, indicating that automated, data-driven parameter selection (e.g., L-curve methods) will be necessary for large-scale clinical application. While the Dual-Tikhonov optimization successfully stabilized the global inversion, it exhibited measurable limitations in accurately preserving the maximum amplitude of the sharp stenotic jet, yielding an optimal RMSE of 7.71 cm/s for Scenario B. This error margin arises from the fundamental mathematical properties of the L2-norm utilized in Tikhonov regularization, which aggressively penalizes steep spatial gradients. Consequently, the optimizer is forced into a mathematical compromise between adequately suppressing high-frequency baseline noise and preserving the true height of sharp hemodynamic transients. While this L2-norm formulation is highly effective for smooth, continuous morphological changes such as distal tapering, it blunts the estimation of severe focal pathologies. To overcome this limitation in future in vivo applications, the inverse solver will be upgraded to incorporate Total Variation (L1-norm) regularization. Unlike the L2-norm, L1 regularization mathematically permits abrupt, discontinuous signal transitions while strictly penalizing oscillatory noise.

Beyond algorithmic regularization, model accuracy is also tied to the scanner and tissue dependent parameters. As illustrated in Figure 3, blood T1 variability and B1+/flip-angle inhomogeneity can affect quantitative velocity estimates because these parameters directly determine $R_{\text{eff}}$, $S_0$, and $S_{\text{ss}}$. The dominant systematic trend was driven by flip-angle scaling, while T1 variability further modified the magnitude of the error. Importantly, the RMSE maps reflect both parameter mismatch and intrinsic inverse-model error, particularly in the stenosis scenario where regularization smooths the sharp velocity peak. Thus, the analysis indicates that the method is relatively stable near nominal assumptions but can exhibit clinically meaningful bias under substantial B1+ or T1 mismatch. Future implementations may benefit from B1+ correction, subject- or field-strength-specific blood T1 assumptions, or joint estimation/marginalization of relaxation and excitation parameters when quantitative velocity accuracy is required.

To visually conceptualize the practical application of this framework, despite these physical variables, we mapped the progressive decrease in macroscopic signal intensity as blood traverses complex, tortuous flowing tubes in vivo (Figure 4A). Utilizing computational tools such as the ArteryX [20] application, we can reliably trace these complex paths and extract the necessary 1D spatial intensity profiles to facilitate the inversion. By applying the Dual-Tikhonov inverse solver to these in vivo topological graphs, we successfully estimated localized velocity (Figure 4B) without the need for supplementary functional imaging. The preliminary results, demonstrated on slab-specific distal arteries, yield velocity estimations that is similar to in vivo ranges reported in the literature [21–23].

However, as this study relies on computational simulations, transitioning to broad in vivo clinical scanning presents real-world challenges stemming from simplified TOF-MRA physics utilized in our model. While global B1+ field inhomogeneities can typically introduce signal bias, our approach inherently mitigates this confound; by proposing to apply the inverse solver piecewise over short, localized vessel segments utilizing ArteryX graph centerline **[20]**, spatial variations in the B1+ field can be rendered negligible within each calculation window. Yet several active limitations remain inherent to this foundational 1D derivation. Primarily, the current framework relies on fixed assumptions regarding the longitudinal relaxation time (T1) of blood, our computational analysis has quantified the model's sensitivity to physiological T1 and B1+ variations. Nonetheless, subsequent in vivo experiments will be needed to clinically validate the approach across diverse, patient-specific T1 blood profiles. explicitly quantify the model's sensitivity to physiological T1 variations and validate the approach across diverse T1 blood profiles. Furthermore, clinical TOF-MRA acquisitions are susceptible to partial-volume effects at vessel boundaries, as well as slab-transition artifacts arising from heterogeneous RF excitation histories in multi-slab imaging. Another important limitation is that TOF-MRA cannot inherently track pulsatile flow dynamics; thus, future work is needed to enable reliable hemodynamic estimation by addressing these systolic and diastolic phase variations.

Additionally, the 1D nature of this model cannot resolve the complex, multi-directional flow characteristic of aneurysms, necessitating future integration with 3D computational fluid dynamics (CFD) to accurately map spatial velocity gradients and directly estimate critical biomarkers such as wall shear stress (WSS). To mitigate spatial artifacts in our preliminary in vivo demonstration, we specifically fitted distal arteries, ensuring they were slab specific. Ultimately, the proposed methodology is expected to perform most robustly in these distal arterial slabs, such as the middle cerebral artery (MCA), because blood resides within these segments for extended durations.

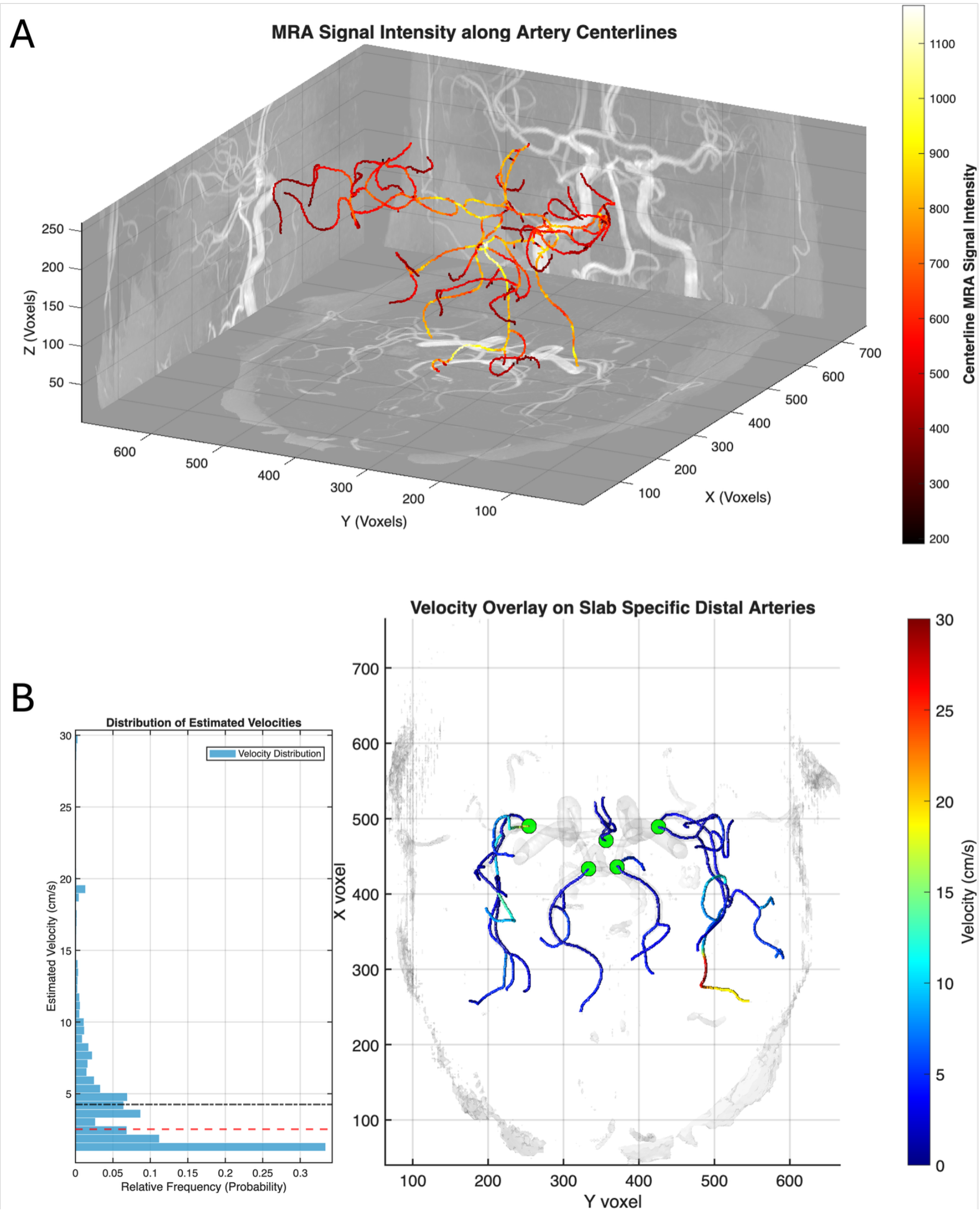

**Figure 4.** (**A**) MATLAB 3D scatter plot overlayed on MIP, visualizing MRA signal intensity along extracted human brain artery centerlines (data from the TopBrain Challenge). Voxel intensity, color-coded according to the vertical scale (right), decreases from the central entry region (yellow) into the distal vascular territories (red/brown). This spatial decay along the intricate centerlines provides a definitive in vivo illustration of the generalized tapering flowing tube geometries simulated in Scenario (**A**). (**B**) Preliminary velocity results demonstrated on slab specific distal arteries; estimation is shown to be within in vivo range from literature which motivates further validation. The black and red dotted lines indicates mean and median respectively.

## 5. Conclusions

This study demonstrates the theoretical and computational feasibility of estimating quantitative, pointwise variable fluid velocity directly from TOF-MRA signal intensity profiles along artery centerlines. By re-evaluating the approach-to-steady-state Bloch equations to account for continuous transit time variations and employing a Dual-Tikhonov global matrix optimization, the proposed framework successfully recovered underlying hemodynamic profiles in both narrowing and stenotic flowing tube models without the need for supplementary functional imaging or multi-echo sequences. While the L2-norm regularization exhibited a minor amplitude attenuation when estimating sharp pathological transients, the inverse solver maintained mathematical stability against high-frequency noise.

With the baseline physical and mathematical framework validated in the simulation, preliminary in vivo results are shown for limited single-slab-specific distal arteries. Future research will translate these methods directly to in vivo clinical datasets to evaluate the topological resilience of the framework against complex, physiological cerebrovascular networks, varying scanner specific parameters, potentially solidifying its utility as a unified structural and functional diagnostic tool.

**Author Contributions:** Conceptualization, A.F.; methodology, A.F.; software, A.F.; validation, A.F.; formal analysis, A.F.; investigation, A.F.; resources, A.F.; data curation, A.F.; writing—original draft preparation, A.F.; writing—review and editing, A.F. and M.N.U. and G.S.; visualization, A.F.; supervision, A.F.; project administration, A.F.; funding acquisition, M.N.U. and G.S. All authors have read and agreed to the published version of the manuscript.

**Funding:** National Institute of Neurological Disorders and Stroke (NINDS) R03NS134395-01 and National Institute of Aging (NIA) R01AG077156-01A1.

**Institutional Review Board Statement:** Not applicable.

**Informed Consent Statement:** Not applicable.

**Data Availability Statement:** The code is made available at https://github.com/abrarfaiyaz/tof_mra_centerline_sim (accessed on 20 August, 2026). Originality: This article represents original work. There is no significant overlap with any other articles by the authors that are under consideration or in press elsewhere except in arxiv. Author Agreement: All co-authors have reviewed and approved the manuscript in its currently submitted form. Open Access: All authors agree that, upon acceptance, this article will be published under a Creative Commons Attribution (CC BY) license. AI Tool Usage: AI tools were used solely to assist with minor language editing and clarity in the drafting process, and all scientific content, methodology, and conclusions are the original work of the authors.

**Conflicts of Interest:** The authors declare no conflicts of interest.

## Abbreviations

The following abbreviations are used in this manuscript:

| | |
|---|---|
| TOF | Time Of Flight |
| MRA | MR Angiography |
| CSVD | Cerebrovascular Small Vessel Disease |
| PC | Phase Contrast |
| DSA | Digital Subtraction Angiography |
| TCD | Transcranial Doppler |
| ASL | Arterial Spin Labeling |
| MCA | Middle Cerebral Artery |
| RMSE | Root Mean Squared Error |
| WSS | Wall Shear Stress |
| MIP | Maximum Intensity Projection |

## References

1. Shi, Z.; Zhao, X.; Zhu, S.; Miao, X.; Zhang, Y.; Han, S.; Wang, B.; Zhang, B.; Ye, X.; Dai, Y.; et al. Time-of-Flight Intracranial MRA at 3 T versus 5 T versus 7 T: Visualization of Distal Small Cerebral Arteries. *Radiology* **2022**, *305*, E72. https://doi.org/10.1148/radiol.229027.
2. Zeng, P.; Zeng, B.; Wang, X.; Yin, F.; Li, B.; Nie, L.; Tian, L.; Luo, D.; Li, Y. Association between carotid artery hemodynamics and neurovascular coupling in cerebral small vessel disease: An exploratory study. *Front. Aging Neurosci.* **2025**, *7*, 17. https://doi.org/10.3389/fnagi.2025.1536552.
3. Leung, J.; Behpour, A.; Sokol, N.; Mohanta, A.; Kassner, A. Assessment of intracranial blood flow velocities using a computer controlled vasoactive stimulus: A comparison between phase contrast magnetic resonance angiography and transcranial doppler ultrasonography. *J. Magn. Reson. Imaging* **2013**, *38*, 733–738. https://doi.org/10.1002/jmri.23911.
4. Lim, R.P.; Shapiro, M.; Wang, E.Y.; Law, M.; Babb, J.S.; Rueff, L.E.; Jacob, J.; Kim, S.; Carson, R.; Mulholland, T.; et al. 3D Time-Resolved MR Angiography (MRA) of the Carotid Arteries with Time-Resolved Imaging with Stochastic Trajectories: Comparison with 3D Contrast-Enhanced Bolus-Chase MRA and 3D Time-of-Flight MRA. *Am. J. Neuroradiol.* **2008**, *29*, 1847–1854. https://doi.org/10.3174/ajnr.A1252. PMID: 18768727.
5. Neumann, K.; Günther, M.; Düzel, E.; Schreiber, S. Microvascular Impairment in Patients With Cerebral Small Vessel Disease Assessed With Arterial Spin Labeling Magnetic Resonance Imaging: A Pilot Study. *Front. Aging Neurosci.* **2022**, *14*, 871612. https://doi.org/10.3389/fnagi.2022.871612.
6. Koktzoglou, I.; Huang, R.; Edelman, R.R. Quantitative time-of-flight MR angiography for simultaneous luminal and hemodynamic evaluation of the intracranial arteries. *Magn. Reson. Med.* **2022**, *87*, 150–162. https://doi.org/10.1002/mrm.28969.
7. Morgan, A.G.; Thrippleton, M.J.; Wardlaw, J.M.; Marshall, I. 4D flow MRI for non-invasive measurement of blood flow in the brain: A systematic review. *J. Cereb. Blood Flow Metab.* **2021**, *41*, 206–218. https://doi.org/10.1177/0271678X20952014.
8. Koktzoglou, I.; Huang, R. Intracranial arterial flow velocity mapping in quantitative time-of-flight MR angiography using deep machine learning. *Magn. Reson. Imaging* **2023**, *100*, 10–17. https://doi.org/10.1016/j.mri.2023.02.005.
9. Nishimura, D. Principles of Magnetic Resonance Imaging. 2010. Available online: https://www.semanticscholar.org/paper/Principles-of-magnetic-resonance-imaging-Nishimura/89d63881af2084f327272030cb59343fca7b919b (accessed on 17 July 2026).
10. Wright, S.N.; Kochunov, P.; Mut, F.; Bergamino, M.; Brown, K.M.; Mazziotta, J.C.; Toga, A.W.; Cebral, J.R.; Ascoli, G.A. Digital reconstruction and morphometric analysis of human brain arterial vasculature from magnetic resonance angiography. *NeuroImage* **2013**, *82*, 170–181. https://doi.org/10.1016/j.neuroimage.2013.05.089.
11. Schnell, S.; Ansari, S.A.; Wu, C.; Garcia, J.; Murphy, I.G.; Rahman, O.A.; Rahsepar, A.A.; Aristova, M.; Collins, J.D.; Carr, J.C.; et al. Accelerated dual-venc 4D flow MRI for neurovascular applications. *J. Magn. Reson. Imaging* **2017**, *46*, 102–114. https://doi.org/10.1002/jmri.25595. PMID: 28152256; PMCID: PMC5464980.
12. Terada, M.; Takehara, Y.; Isoda, H.; Wakayama, T.; Nozaki, A. Technical Background for 4D Flow MR Imaging. *Magn. Reson. Med. Sci.* **2022**, *21*, 267–277. https://doi.org/10.2463/mrms.rev.2021-0104.
13. Liu, J.; Koskas, L.; Faraji, F.; Kao, E.; Wang, Y.; Haraldsson, H.; Kefayati, S.; Zhu, C.; Ahn, S.; Laub, G.; et al. Highly Accelerated Intracranial 4D Flow MRI: Evaluation on Healthy Volunteers and Patients with Intracranial Aneurysms. *Magn. Reson. Mater. Phys. Biol. Med.* **2018**, *31*, 295–307. https://doi.org/10.1007/s10334-017-0646-8. PMID: 28785850; PMCID: PMC5803461.
14. Purkayastha, S.; Sorond, F. Transcranial Doppler Ultrasound: Technique and Application. *Semin. Neurol.* **2012**, *32*, 411–420. https://doi.org/10.1055/s-0032-1331812.
15. Pan, Y.; Wan, W.; Xiang, M.; Guan, Y. Transcranial Doppler Ultrasonography as a Diagnostic Tool for Cerebrovascular Disorders. *Front. Hum. Neurosci.* **2022**, *16*, 841809. https://doi.org/10.3389/fnhum.2022.841809.
16. Haller, S.; Zaharchuk, G.; Thomas, D.L.; Lovblad, K.O.; Barkhof, F.; Golay, X. Arterial Spin Labeling Perfusion of the Brain: Emerging Clinical Applications. *Radiology* **2016**, *281*, 337–356. https://doi.org/10.1148/radiol.2016150789.
17. Alsop, D.C.; Detre, J.A.; Golay, X.; Günther, M.; Hendrikse, J.; Hernandez-Garcia, L.; Lu, H.; MacIntosh, B.; Parkes, L.M.; Smits, M.; et al. Recommended implementation of arterial spin-labeled perfusion MRI for clinical applications: A consensus of the ISMRM perfusion study group and the European consortium for ASL in dementia. *Magn. Reson. Med.* **2015**, *73*, 102–116. https://doi.org/10.1002/mrm.25197.
18. Nam, H.H.; Jang, D.K.; Cho, B.R. Complications and risk factors after digital subtraction angiography: 1-year single-center study. *J. Cerebrovasc. Endovasc. Neurosurg.* **2022**, *24*, 335–340. https://doi.org/10.7461/jcen.2022.E2022.05.001.

19. Huang, T.C.; Chang, C.K.; Liao, C.H.; Ho, Y.J. Quantification of Blood Flow in Internal Cerebral Artery by Optical Flow Method on Digital Subtraction Angiography in Comparison with Time-of-Flight Magnetic Resonance Angiography. *PLoS ONE* **2013**, *8*, e54678. https://doi.org/10.1371/journal.pone.0054678.
20. Faiyaz, A.; Hoang, N.; Schifitto, G.; Uddin, M.N. ArteryX: A Reliable End-to-End Toolbox for Standardized Intracranial Artery Feature Extraction from 3D TOF-MRA. *Magn. Reson. Med.* **2026, 1-11**. https://doi.org/10.1002/mrm.70564.
21. Zhou, S.; Gao, X.; Park, G.; Yang, X.; Qi, B.; Lin, M.; Huang, H.; Bian, Y.; Hu, H.; Chen, X.; et al. Transcranial volumetric imaging using a conformal ultrasound patch. *Nature* **2024**, *629*, 810–818. https://doi.org/10.1038/s41586-024-07381-5.
22. Schnell, S.; Ansari, S.A.; Vakil, P.; Wasielewski, M.; Carr, M.L.; Hurley, M.C.; Bendok, B.R.; Batjer, H.; Carroll, T.J.; Carr, J.; et al. Three-dimensional hemodynamics in intracranial aneurysms: Influence of size and morphology. *J. Magn. Reson. Imaging* **2014**, *39*, 120–131. https://doi.org/10.1002/jmri.24110.
23. Bisbal, J.; Winter, P.; Jofre, S.; Ponce, A.; Ansari, S.A.; Abdalla, R.; Markl, M.; Odeback, O.W.; Uribe, S.; Tejos, C.; et al. Uncertainty-aware automated labeling of intracranial arteries using deep learning. *BMC Med. Imaging* **2026**, *26*, 175. https://doi.org/10.1186/s12880-026-02276-5. PMID: 41840364; PMCID: PMC13045049.

**Disclaimer/Publisher's Note:** The statements, opinions and data contained in all publications are solely those of the individual author(s) and contributor(s) and not of MDPI and/or the editor(s). MDPI and/or the editor(s) disclaim responsibility for any injury to people or property resulting from any ideas, methods, instructions or products referred to in the content.